\documentclass[11pt,letterpaper]{article}
\usepackage[letterpaper,bindingoffset=0.2in,%
            left=1in,right=1in,top=1in,bottom=1in,%
            footskip=.25in]{geometry}
            
\usepackage{amsmath}
\usepackage{amssymb}

\usepackage[hidelinks]{hyperref}

\usepackage{lineno}

\usepackage{color}

\usepackage{microtype}
\DisableLigatures[f]{encoding = *, family = * }

\usepackage{setspace}

\usepackage{graphicx}

\usepackage{lineno}
\usepackage[font=small,format=plain,labelfont=bf,up,textfont=up]{caption}

\usepackage{abstract}

\begin{document}


\begin{center}
{\Large
\textbf{\textcolor{black}{
Evidence for a multi-level trophic organization \\of the human gut microbiome 
}}
}
\\
\vskip 10pt 
\normalsize
Tong Wang \textsuperscript{1, 2,$\dagger$},
Akshit Goyal \textsuperscript{3, $\dagger$},
Veronika Dubinkina \textsuperscript{2, 4},
Sergei Maslov \textsuperscript{2, 4, *}
\\
\bigskip
\textbf{1} Department of Physics,
University of Illinois at Urbana-Champaign, IL  61801, USA
\\
\textbf{2} Carl R. Woese Institute for Genomic Biology, 
University of Illinois at Urbana-Champaign, IL  61801, USA
\\
\textbf{3} Simons Centre for the Study of Living Machines, 
National Centre for Biological Sciences, 
Tata Institute of Fundamental Research, 
Bengaluru 560 065, India
\\
\textbf{4} Department of Bioengineering, 
University of Illinois at Urbana-Champaign, IL  61801, USA
\\
\bigskip

\bigskip

%
%
\vskip 10pt
\small
$\dagger$ These authors contributed equally to this work.
\\
$\ast$ Correspondence: \href{mailto:maslov@illinois.edu}{\texttt{maslov@illinois.edu}}
\end{center}

\section*{Abstract}
\noindent
The human gut microbiome is a complex ecosystem, in which hundreds of microbial species and metabolites coexist, in part due to an extensive network of cross-feeding interactions. However, both the large-scale trophic organization of this ecosystem, and its effects on the underlying metabolic flow, remain unexplored. Here, using a simplified model, we provide quantitative support for a multi-level trophic organization of the human gut microbiome, where microbes consume and secrete metabolites in multiple iterative steps. Using a manually-curated set of metabolic interactions between microbes, our model suggests about four trophic levels, each characterized by a high level-to-level metabolic transfer of byproducts. It also quantitatively predicts the typical metabolic environment of the gut (fecal metabolome) in approximate agreement with the real data. To understand the consequences of this trophic organization, we quantify the metabolic flow and biomass distribution, and explore patterns of microbial and metabolic diversity in different levels. The hierarchical trophic organization suggested by our model can help mechanistically establish causal links between the abundances of microbes and metabolites in the human gut.

\section*{Introduction}
The human gut microbiome is a complex ecosystem with several hundreds of microbial species \cite{HMP2012, Qin2010} consuming, producing and exchanging hundreds of metabolites 
\cite{magnusdottir2017generation, san2018enormous, pacheco2019costless, sung2017global, goyal2018metabolic}. 
With the advent of high-throughput genomics and metabolomics techniques, it is now possible to simultaneously measure the levels of individual metabolites (the fecal metabolome), as well as the abundances of individual microbial species \cite{ponomarova2015metabolic}. 
Quantitatively connecting these levels with each other, requires knowledge of the relationships between microbes and metabolites in their shared 
environment: who produces what, and who consumes what? \cite{muller2018using, bauer2018network}
In recent studies, information about these relationships for all of the common species and metabolites in the human gut 
has been gathered using both manual curation from published studies \cite{sung2017global} and automated genome 
reconstruction methods \cite{magnusdottir2017generation}. This has laid the foundation for 
mechanistic models which would allow one to relate metabolome composition to microbiome composition 
\cite{magnusdottir2018modeling, garza2018towards}. 

%
%
More generally, the construction of mechanistic models has been hindered by the complexity of dynamical processes taking place in the human gut, which in addition to 
cross-feeding and competition, includes differential spatial distribution and species motility, interactions of microbes 
with host immune system and bacteriophages, changes in activity of metabolic pathways in individual species in response to 
environmental parameters, etc. This complexity can be tackled on several distinct levels. For 2-3 species it is possible to construct a detailed dynamical model taking into account the spatial organization and flow of microbes and nutrients within the lower gut \cite{cremer2016effect, cremer2017effect}, or optimizing the intracellular metabolic flows as well as competition for extracellular nutrients using dynamic flux balance analysis (dFBA) models \cite{varma1994stoichiometric, san2018enormous}.

For around 10 microbial species, and a comparable number of metabolites, 
it is possible to construct a consumer-resource model (CRM) taking into account microbial competition for nutrients \textcolor{black}{\cite{waltman2017theoretical}}, the generation of metabolic byproducts \textcolor{black}{\cite{goldford2018emergent}}, and the different tolerance of species to various environmental factors like pH \textcolor{black}{\cite{cremer2017effect, ratzke2018modifying}}. Using the existing experimental data on consumption and production kinetics of different metabolites, it is possible to fit some (but not all) of around 80 parameters in such a model \cite{kettle2015modelling}. \textcolor{black}{These models are also capable of incorporating cross-feeding interactions between microbial species, as well as community assembly processes \cite{kettle2015modelling, louca2016reaction}.}

However, modeling 100s of species and metabolites, typically present in an individual's gut microbiome, requires thousands of parameters, which cannot be estimated from the current experimental data. Therefore, any such model must instead resort to a few ``global parameters'' that appropriately coarse-grain the relevant ecosystem dynamics. Here, we propose such a coarse-grained model of the human gut microbiome, hierarchically organized into several distinct trophic levels. In each level, metabolites are consumed by a subset of microbial species in the microbiome, and partially converted to microbial biomass. A remainder of these metabolites is excreted as metabolic byproducts, which then form the next level of metabolites. The metabolites in this level can then be consumed as nutrients by another subset of microbial species. Our model needs two global parameters: (1) the fraction of nutrients converted to metabolic byproducts by any microbial species, and (2) the number of trophic levels into which the ecosystem is hierarchically organized. 

While previous studies have suggested that such cross-feeding of metabolic byproducts is common in the microbiome, the extent to which this ecosystem is hierarchically organized has not been quantified.
Our model suggests that both, the gut microbiome, and its relevant metabolites, are organized into roughly 4 trophic levels, which interconnect these microbes and metabolites in quantitative agreement with their experimentally measured levels. We also show that this model can predict the flow of biomass and metabolites through these trophic levels, quantify the relative contribution of the observed microbes and metabolites to these levels, and \textcolor{black}{thereby describe the effective diversity} at each level. 



\section*{Model and Results}
\subsection*{Multi-level trophic model of the human gut microbiome}
\begin{figure}[!t]
\includegraphics[width=\linewidth]{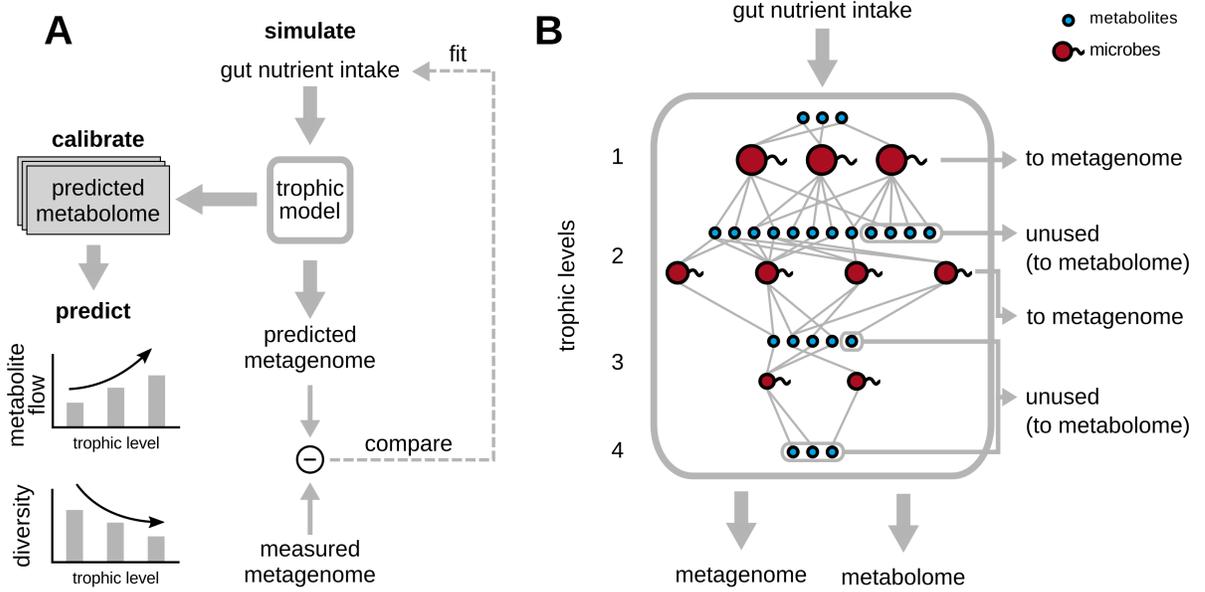}
\caption{{\bf Overview of the trophic model, its calibration and predictions.}
(A) Schematic diagram showing the various steps in the trophic model, which uses fits the gut nutrient intake profile best approximating the measured metagenome, and outputs a predicted metagenome (microbial abundances) and metabolome. 
The experimentally measured metabolome is used to calibrate the number of trophic levels, $N_\ell$ and byproduct fraction, $f$ of the model. 
(B) ``Zoomed-in'' view of the trophic model from (A), with different microbial species (red) and metabolites (blue) spread across the four trophic levels suggested by the model. At each level, metabolites are consumed by microbial species, and converted partially to their biomass, while the remainder is secreted as metabolic byproducts, which are nutrients for the next trophic level. Metabolites that are left unconsumed across each level are assumed to eventually exit the gut as part of the fecal metabolome, while the biomass accumulated by each species across all levels contributes to the metagenome.
}
\label{fig1}
\end{figure}

Our model aims to approximate the metabolic flow through the intricate cross-feeding network of microbes in the lower intestine (hereafter, ``gut'') human individuals (figure \ref{fig1}A). This flow begins with metabolites entering the gut, which are subsequently consumed and processed by multiple microbial species. We assume that each microbial species grows by converting a certain fraction of its metabolic inputs (nutrients) to its biomass
and secretes the rest as metabolic byproducts (figure \ref{fig1}B). We define the byproduct fraction, $f$, one of the two key parameters of our model, as the fraction of nutrients secreted as byproducts. The complementary biomass fraction, $1-f$, is the fraction of nutrient inputs converted to microbial biomass. 
The metabolic byproducts produced from the nutrients entering the gut, can be further consumed by some species in the microbiome, in turn generating a set of secondary metabolic byproducts. We call each step of this process of metabolite consumption and byproduct generation, a trophic level. Due to factors such as limited gut motility, and a finite length of the lower gut, we assume that this process only continues for a finite 
number of levels, $N_\ell$, the second key parameter of our model. At the end of this process, metabolites left unconsumed after 
passing through $N_\ell$ trophic levels 
are assumed to leave the gut as a part of the feces (figure \ref{fig1}B).

In order to quantitatively describe all the steps of this process, our model requires the following information: 
\begin{itemize}
    \item The metabolic capabilities of different microbial species in the gut, i.e., which microbes can consume which metabolites, and secrete which others. 
    For this, we used a manually curated database connecting 567 common human gut microbes to 235 gut-relevant metabolites 
    they are capable of either consuming or producing as byproducts \cite{sung2017global} (see Methods for details).
    \item The nutrient intake to the gut, which is the first set of metabolites that are consumed by the microbiome. Since the levels of these metabolites in a given individual are generally unknown, we first curated a list of 19 metabolites likely to constitute the bulk of this nutrient intake, and subsequently fitted their levels to best describe the observed microbial abundances in the gut of each individual (see Methods). We collected such microbial abundance data from various sources, in particular: 380 samples  from the large-scale whole-genome sequencing (WGS) studies of healthy individuals (Human Microbiome Project (HMP)  \cite{HMP2012} and the MetaHIT consortium \cite{Qin2010, qin2012metagenome}), 41 samples from a recent 16S rRNA 
    study of 10 year old children in Thailand \cite{kisuse2018urban}.
    \item The kinetics of nutrient uptake and byproduct release, i.e., the rates we refer to as $\lambda$'s, at which different microbial species obtain and secrete different metabolites in the gut environment {\color{black}(see Methods for details of how we defined $\lambda$'s)}. Since this information is unknown for most of our microbes and metabolites, we made some simplifying assumptions. We assumed that, in a given level, when species consume the same metabolite, 
    they receive it in proportion to their 
    abundance in the microbiome. When secreting metabolic byproducts, we assumed equal splitting, such that every metabolite secreted by a given species was released in the same fraction. However, we later verified that the predictions of our model was relatively insensitive to the exact values of these parameters, by repeating our simulations with randomized values of these parameters (see figure S1).
\end{itemize}



%
%
{\color{black}
\subsection*{Simulating the trophic model}


Our model describes the transit of nutrients from the lower gut to the feces of a specific human individual. As the nutrients transit through the gut, the microbial species in the gut consume, digest and convert them to microbial biomass and metabolic byproducts.
For  a specific individual, our model comprises multiple iterative steps of metabolite consumption by microbes and the subsequent generation of metabolic byproducts, with each step constituting a trophic level. At each level, all metabolites produced in the previous level could be consumed by all microbial species detected in the specific individual's gut. Note that at the first level, these metabolites were given by the nutrient intake to the gut, as described above. Any metabolite that could be consumed by multiple microbial species, was split across those species in proportion to their experimentally measured relative abundances (see Methods for details). 
Those metabolites that could not be consumed at any level were assumed to eventually exit the gut, and form part of the individual's fecal metabolome. Upon metabolite consumption in any trophic level, we assumed that all microbial species that consumed these metabolites and converted a fraction $(1-f)$ of the total consumed metabolites to their biomass. The remaining fraction, $f$ (assumed fixed for all species) was converted to byproducts for the next level. Here, we assumed that each of the species produced all the byproducts it was capable of in equal amounts. After $N_\ell$ such iterative rounds (calibrated separately, see the next section), we assumed that this process ends. We added up all the biomass accumulated by each microbial species across all trophic levels as their total biomass, and added up all the unconsumed metabolite levels as the total fecal metabolome. Finally, we normalized, both the microbial biomass and metabolite amounts separately, to obtain the relative microbial abundances and relative metabolome profiles, respectively.
}

\subsection*{Calibrating the key parameters of the model}
\begin{figure}[!t]
\centering
\includegraphics[width=0.8\linewidth]{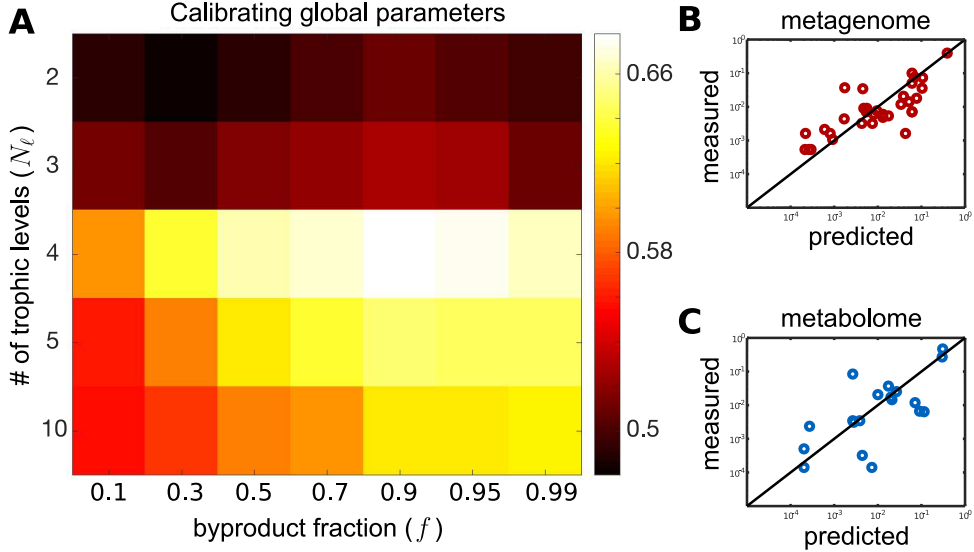}
\caption{{\bf Calibration of the model.}
(A) Heatmap of the Pearson correlation between experimentally measured and predicted metabolomes for different combinations of parameters $f$ and $N_\ell$. The plotted value is the correlation coefficient averaged over 41 individuals in Ref. \cite{kisuse2018urban} (B) Comparison between 
the experimentally observed bacterial abundances in a representative individual (y-axis) and their best fits from our model (x-axis) with 
$f=0.9$ and $N_\ell=4$. (C) Comparison between the experimentally observed fecal metabolome (y-axis) and the predictions of our model (x-axis) with 
$f=0.9$ and $N_\ell=4$ in the same individual shown in panel (B) (Pearson correlation coefficient $0.68$; $P$-value $< 10^{-5}$).}
\label{fig2}
\end{figure}

To calibrate the two key parameters of our model, $f$ and $N_\ell$, we used data from the 41 individuals from a recent 16S rRNA sequencing study of Thai children \cite{kisuse2018urban} for which both, 16S rRNA metagenomic profiles, as well as quantitative levels of 214 metabolites in the fecal metabolome, were available. {\color{black} We used these data specifically because they had simultaneously measured the metagenomes and fecal metabolomes with high accuracy, i.e., at the level of individual species and metabolites, which we required for calibration.}
In each individual we fitted the nutrient intakes of the 19 metabolites to best agree with experimental microbial abundances. 
A representative example comparing the predicted and measured bacterial abundances is shown in figure \ref{fig2}B. 
The Pearson correlation coefficient for data shown in this plot is $0.94$, while in individual participants it ranged between $0.81 \pm 0.17$. 

We carried out these fits of microbial abundances for each of the 41 individuals studied in Ref. \cite{kisuse2018urban} for a broad range of two parameters of our model - the byproduct fraction $f$ ranging between $0.1$ and $0.99$ and the number of trophic levels $N_\ell$ between $2$ and $10$. For each individual and each pair of parameters $f$ and $N_\ell$ we used our model to predict the fecal metabolome profile. This predicted metabolome was subsequently compared to the experimental data of Ref. \cite{kisuse2018urban} measured in the same individual. Around 19 of our predicted metabolites (variable across individuals) were actually among the ones experimentally measured in Ref. \cite{kisuse2018urban}. 
{\color{black}We quantified the quality of our predictions using the Pearson correlation coefficient between the predicted and experimentally measured metabolomes, and it's associated $P$-value. 
The model with parameters $f=0.9$ and $N_\ell=4$ best agreed with the experimental metabolome data, among all the values we tried (Pearson correlation $0.7 \pm 0.2$; median $P$-value $8 \times 10^{-4}$; see figure \ref{fig2}A). To account for the fact that we used two adjustable parameters in our model ($f$ and $N_\ell$, we have corrected the $P$-values appropriately (see Methods for details). We found that even after this correction the median $P$-value $\sim 10^{-3}$ is well below the commonly used significance threshold of $0.05$. 
To ensure that our calibration was not sensitive to this specific measure of fit quality, we also calculated an alternative measure --- that of a logarithmic accuracy --- which quantifies the average order-of-magnitude error in our predicted fecal metabolome, when compared with the experimentally measured one (see Methods for details). We found that the best logarithmic accuracy was still achieved in a model with $f=0.9$ and $N_\ell=4$ (the mean error is $0.8$ orders of magnitude; see figure \ref{suppfig4}). Hence, we used this combination of parameters in all subsequent simulations of our model.}

An example of the agreement between predicted and experimentally observed fecal metabolome in a single individual (the same one as in figure \ref{fig2}B) is shown in 
figure \ref{fig2}C ({\color{black}Pearson correlation coefficient 0.89; the adjusted $P$-value $< 10^{-6}$)}. 
Note that, while the agreement between the experimentally observed and predicted microbial abundances shown in figure \ref{fig2}B 
is the outcome of our fitting the levels of 19 intake metabolites, the fecal metabolome is an independent prediction 
of our model. It naturally emerges from the trophic organization of the metabolic flow and agrees well with the 
experimentally observed metabolome. {\color{black}To test the quality of this independent prediction, and to show its dependence on metabolic interactions, we repeated our simulations using a randomly shuffled set of microbial metabolic capabilities (i.e., we independently shuffled consumption and secretion abilities of individual microbial species; see Methods for details). Figure \ref{suppfig3} shows the model results generated by this shuffled microbial metabolic capabilities. We found that the model now generated a much worse correlation coefficient, and more importantly, a non-significant median $P$-value 0.05 which did not clear the commonly used threshold of $P < 0.05$ (for example, the individual in figure \ref{fig2}B--C has Pearson correlation 0.32; $P$-value $=0.19$; see figure \ref{suppfig3}). For all individuals, the Pearson correlation is $0.44 \pm 0.2$ and the median of their corresponded $P$-value $0.046$. Taken together, our simplified model
supports the organization of the human gut microbiome into roughly four trophic levels with byproduct fraction around 0.9.
} 

{\color{black} To apply our model to broader, more representative and better-studied samples of the human gut microbiome, we carried over the results of this calibration to another dataset. This dataset (discussed in the next section) consisted of a cohort of 380 human individuals from the Human Microbiome Project (HMP) and the MetaHIT study. We carried over this calibration for three reasons: (1) the lack of availability of simultaneous metabolome measurements for the latter dataset; (2) the fact that both datasets are derived from the human gut; and (3) the similarity in the level of metagenome variability in both datasets.}





\subsection*{Predictions of the multi-level trophic model}
\subsubsection*{Metabolite and biomass flow through trophic levels}
\begin{figure}[!t]
\centering
\includegraphics[width=0.8\linewidth]{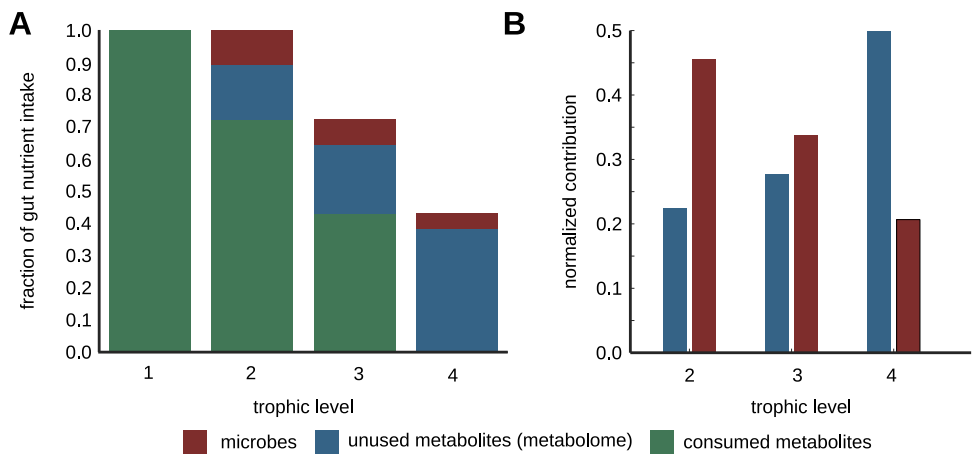}
\caption{{\bf Metabolite and biomass flow through the levels.}
(A) Cascading nature of nutrient flow across trophic levels: nutrient intake to the gut (the leftmost turquoise bar) is gradually 
converted into microbial biomass (red bars in each level) and metabolic byproducts (turquoise bars in each level). 
Some fraction of these byproducts (blue bars in each level) cannot be consumed by the microbiome and hence remains further unprocessed 
until it leaves an individual as their fecal metabolome. The metabolic byproducts of each level (turquoise bars) serve as the nutrient 
intake for microbes in the next level. The process ends at level 4 where all byproducts remain unconsumed thereby 
enter the fecal metabolome. (B) Normalized contribution of of the nutrient intake to microbial biomass (red) and fecal 
metabolome (blue) split across levels 2 to 4.
Dashed lines show that consumable metabolites generated at a previous level serve as metabolic inputs to the next level.}
\label{fig3}
\end{figure}
With a well-calibrated and tested model we are now in a position to apply it to a broader set of human microbiome data. 
To this end we chose data for 380 healthy adult individuals from several countries (Europe \cite{Qin2010}, USA \cite{HMP2012}, and China \cite{qin2012metagenome}). For each individual, we used our model to predict its metabolome (that has not been measured experimentally) and quantified the flow of nutrients (or metabolic activity) through 4 trophic levels in our model averaged over these individuals. 

Figure \ref{fig3}A shows the cascading nature of this flow: metabolites enter the gut as nutrient intake shown as the leftmost turquoise bar in figure \ref{fig3}A. 
Roughly, a fraction $1-f=0.1$ of this nutrient intake is converted into microbial biomass (red bar), while the remaining fraction $f=0.9$ is excreted as metabolic byproducts. Some fraction of these metabolic byproducts (blue bar) cannot be consumed by any of the microbes in individuals microbiome and hence ultimately it leaves the individual as part of their fecal metabolome. The metabolic byproducts that can be consumed by the microbiome (turquoise bar) serve as the nutrient intake for microbes in the next level (i.e., level 3). This scenario repeats itself over the next levels until the level 4, beyond which we assume all the byproducts enter the fecal metabolome. Note that, even though some of these byproducts can be consumed by gut microbes, our previous calibration (figure \ref{fig2}A) suggests that this does not happen. We believe this may be due to the finite time of flow of nutrients through the gut.
Figure \ref{fig3}B shows the normalized contributions of the nutrient intake to microbial biomass (red) and fecal metabolome (blue) split across trophic levels. We observe a contrasting pattern across levels, with the contribution to microbial biomass decreasing along levels, whereas the fraction of unused metabolites (contribution to the fecal metabolome) increases.
It is also worth noting that the same microbial and metabolic species get contributions from 
multiple trophic levels, i.e., the same microbes that consume nutrients and excrete byproducts in earlier 
levels can also grow on metabolites generated in later levels. Thus, even though the dominant contribution 
to a species' biomass is typically derived from a specific trophic level, species can grow by consuming 
metabolites from multiple levels. 

\begin{figure}[t!]
\centering
\includegraphics[width=0.8\linewidth]{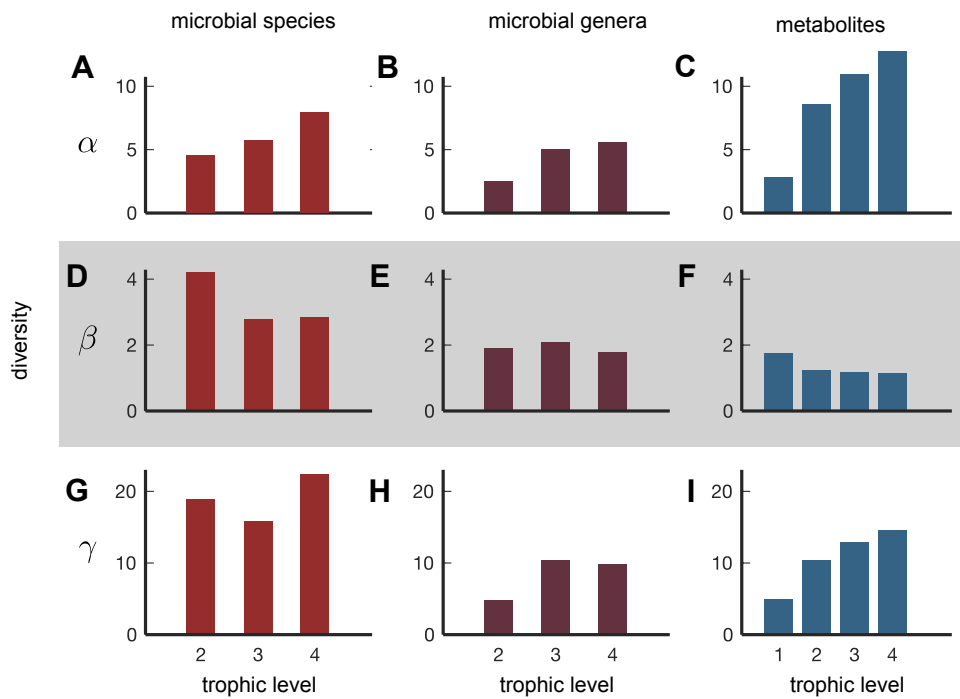}
\caption{{\bf Metabolite and microbial diversity at different levels.}
Effective (A--C) $\alpha$-diversity, (D--F) $\beta$-diversity, (G--I) $\gamma$-diversity in microbial species (A, D, G), microbial genera 
(B, E, H), and metabolites (C, F, I) plotted as a function of trophic level (1--4) and averaged across 380 individuals. }
\label{fig4}
\end{figure}

\subsubsection*{Quantifying diversity across trophic levels}
The diversity of microbial communities can be separately defined both phylogenetically and functionally. Phylogenetic diversity counts the number of abundant microbial species inferred from the metagenomic profile. On the other hand, functional diversity quantifies the variety of collective metabolic activities of these species, which in our case could be inferred from the metabolome profile. Our model allows to quantify both types of diversity on a level-by-level basis. 
Instead of just calculating the presence or absence of microbial species or metabolites at each level, we weighed each microbe or metabolite by their relative contribution to the metabolic activity at that trophic level. 
At each level, we calculated the effective $\alpha$-, $\beta$- and $\gamma$-diversity, separately 
for microbes and metabolites
(see Methods for details).

Figure \ref{fig4} shows the effective $\alpha$-, $\beta$- and $\gamma$- diversity for microbes (grouped at the species and genus levels) and metabolites, averaged over our 380 healthy individuals.
The microbes first appear in the second trophic level feeding off the nutrient intake metabolites in the first level.
We found that the $\alpha$-diversity (the average number of abundant entities  weighted by their contribution to each level) systematically 
increases with the level number for both microbes and metabolites. There is no clear trend in 
the $\gamma$-diversity of microbes grouped at the species level
(the ``pan-microbiome'' diversity, i.e., the number of abundant species in the combined metagenomes of 380 individuals). 

Finally the beta-diversity of microbial species, defined as the ratio between $\gamma-$ and $\alpha$-diversity is the highest ($\sim 4$) in the first level, while being considerably lower ($\sim 2.5$) in the next two levels. The $\beta$-diversity addresses the following important question: how variable are the abundant species between individuals? 

While we found that the $\beta$-diversity of microbial species could be as large as $4$ (figure \ref{fig4}), when we grouped organisms by their genus, $\beta$ diversity decreased down to $\sim 2$ across all levels (figure \ref{fig4}E). This drop in $\beta$-diversity was the most pronounced in the uppermost trophic level. The overall reduction of $\beta$-diversity shown in figure \ref{fig4}E relative to figure \ref{fig4}D suggests that the chief driver of species variability in the gut microbiome is within-genus competition. Such a pattern has previously been explained by a ``lottery-like'' process of microbial competition within the gut \cite{verster2018competitive}.

We also quantified the diversity of metabolites across 4 trophic levels. We found that the $\beta$ diversity of metabolites was the highest in the uppermost level of nutrients ($\sim 2$) and lower in the next three levels ($\sim 1$). While this declining trend was similar to that observed for microbial diversity, surprisingly, the value of $\beta$ diversity for nutrients was much smaller than for microbes (about 2.5 times lower across all levels). This suggests the picture of functional stability --- in spite of taxonomic variability --- in all trophic levels of the human gut microbiome, namely that even though the species composition of the microbiome can be quite different for different individuals, their metabolic function is quite similar. These results supplement similar findings of the HMP project \cite{HMP2012} by breaking them up into trophic levels and by using metabolome diversity instead of metabolic pathways diversity to quantify the extent of functional similarity.


\section*{Discussion}

Above we introduced and studied a mechanistic, consumer-resource model of the human gut microbiome quantifying the flow of metabolites and the gradual building up of microbial biomass across several trophic levels.  What distinguishes our model is its ability to simultaneously capture the metabolic activities of hundreds of species consuming and producing hundreds of metabolites. Using only the metabolic capabilities --- who eats what, and makes what --- of different species in the microbiome, we uncovered roughly four trophic levels in the human gut microbiome. At each of these levels, some microbes consume nutrients, and convert them partially to their biomass, while the remainder gets secreted as metabolic byproducts. These metabolic byproducts can then serve as nutrients for microbes in the next trophic level. 

Understanding such a trophic organization of microbial ecosystems is important because it helps identify causal relationships between microbes and metabolites at two consecutive trophic levels and helps to separate them from purely correlative connections, either at the same or at more distant levels. Thus it extends the previously introduced concept of a ``microbial metabolic influence network'' \cite{sung2017global} by highlighting its hierarchical structure in which species/metabolites assigned to higher trophic levels could affect a large number of species/metabolites located downstream from them. 

{\color{black} The concept of trophic levels has been widely 
used in macro-ecosystems to make sense of flow of nutrients and energy in large food webs \cite{voigt2003trophic, pace1999trophic, williams2004limits}, but it has only received limited attention in microbial ecosystems, one example being ref. \cite{solden2018interspecies}. 
There is no absolute consensus definition of a trophic level with several interpretations  discussed in Refs.
\cite{pimm1977number, burns1989lindeman, kozlovsky1968critical}. 
However, all of these definitions agree with each other on the following two criteria that the trophic structure of an ecosystem typically satisfies: (1) there is explicit level-to-level conversion and flow of energy (and biomass), taking place in several discrete steps; and (2) these steps are temporally staged, 
because the conversion process at every level takes a finite amount of time. Here we define a trophic level as a discrete step in the 
metabolic conversion of nutrients after it enters the lower gut. 
Each such step involves 
multiple microbial species generating byproducts for the next conversion 
step. Thus, according to our definition, the same species and metabolite 
can be present at more than one trophic level. Furthermore, because of 
the finite motility in the human gut, the metabolic 
activity at each of our trophic levels would 
tend to be spatially separated with that in level 1 taking place near 
the entrance to the lower gut and that in level 4, near the end of the gut.
This definition of trophic levels also results in an imperfect hierarchical structure of the food web in which some species or metabolites linking non-consecutive trophic levels (see Ref. \cite{burns1989lindeman} for similar processes in macroscopic ecosystems). Also note that spatially separated microbial compositions, corresponding to the trophic levels in our model, could in principle be tested in artificial gut systems (such as in Refs. \cite{van2015simulator, barroso2015development}).

Further, there are several well-known ecological factors that constrain the number of trophic levels in an ecosystem, such as ecological energetics and population dynamics (see ref. \cite{pimm1977number} for a discussion). Our work introduces additional factors that can limit the number of trophic levels in the human gut microbiome --- namely the limited length and finite motility of the gut.
}

 {\color{black}
The human gut microbiome is notorious for several complex and interlinked metabolic cross-feeding interactions between its resident microbial species \cite{sung2017global, blaut2011ecology}. Even though we exploit this aspect of the gut's microbial ecology to study its trophic organization, we wish to highlight that we do not confine a metabolite or microbial species to participate strictly at one trophic level. We can nevertheless tentatively assign metabolites and microbial species to the level to which they contribute the most. We find that doing so results in trophic level assignments that are consistent with the expectations of the rest of the gut microbial literature \cite{louis2014gut}; see figure \ref{suppfig6} for a representative example of a trophic network. Specifically, we find that various polysaccharide-degrading species from the genera \textit{Prevotella} and \textit{Bacteroides} tend to be assigned to the first microbial layer, leading to the production of acetate \cite{louis2014gut}. This acetate is, in turn, the major substrate for butyrate-producing bacteria such as various species of \textit{Eubacterium} and \textit{Roseburia}, as well as the well-known \textit{Faecalibacterium prausnitzii}; our tentative assignment procedure places these species in the subsequent layers of the trophic network. The butyrate and valerate secreted by these species consequently end up, and are assigned to, metabolite trophic levels 3 and 4. Similarly, various sulfate-reducing species (e.g., \textit{Desulfovibrio piger}, \textit{Bilophila wadsworthia}) and acetogenic bacteria (e.g., \textit{Blautia hansenii}), as well as their byproducts, are typically assigned to the lower trophic levels by our model. One can also see that, towards the lower trophic levels metabolites are either very simple and energy-poor, like CO$_2$, H$_2$, H$_2$S, or are those that cannot be consumed by any gut microbial species, such as various amines, short-chain fatty acids (SCFAs), and secondary bile acids. We expect these latter set of metabolites to therefore be present in an individual's fecal metabolome.
}

\textcolor{black}{By assuming such a fluid multi-level trophic organization, our model is able to independently predict the fecal metabolome of individual humans, in quantitative agreement with experimental measurements, comparable to or better than the state of the art. For example, Ref. \cite{garza2018towards} used intra-cellular metabolic flux balance analysis (FBA) to achieve a Pearson correlation coefficient 0.4 between the predicted and a representative experimentally measured fecal metabolome. In contrast, our model achieved the Pearson correlation of 0.68 in individualized predictions using only two ecologically meaningful parameters. This suggests that incorporating ecological information about the human gut microbiome can generate mechanistically-grounded and internally consistent fecal metabolome predictions given information about an individual's metagenome (species abundance profile).}


Our model also allows us to quantify the diversity of both species and metabolites contributing to different trophic levels. One conclusion we made was that the functional convergence of the microbiome holds roughly equally across all trophic levels. Indeed, at each level we observed the microbial diversity across different individuals was considerably higher than their metabolic diversity. 
Our model also provides additional support to the ``lottery''  scenario described in Ref. \cite{verster2018competitive}, especially in the first trophic level. According to this scenario, there are multiple species nearly equally capable of occupying a certain ecological niche, which in our model corresponds to the set of nutrients they consume and secrete as byproducts. The first species to occupy this niche prevents equivalent microbes from entering it. This is reflected in a high $\beta$-diversity of microbial species combined with a low to moderate $\beta$-diversity of microbial genera to which they belong and low $\beta$-diversity of their metabolic byproducts.

Our model is focused on studying the effects of cross-feeding and competition of different microbes for their nutrients. Thereby it ignores a number of important factors known to impact the composition of the human gut microbiome. These include interactions with host and its immune system \cite{nicholson2012host} as well as with viruses \cite{manrique2016healthy}, and environmental parameters other than nutrients, such as pH \cite{cremer2017effect}, spatial organization \cite{tropini2017gut}, etc. 
Instead, our model uses only two adjustable parameters: the byproduct fraction $f$ and the number of trophic levels $N_\ell$, assumed to be common to all species. This very small number of parameters has been a conscious choice on our part. We are perfectly aware that species differ from each other in their byproduct ratios, and that the metabolic flows are not equally split among multiple byproducts.
This can be easily captured by a variant of our model in which different nutrient inputs and and byproduct outputs of a given microbial species are characterized by different kinetic rates. However, this would immediately increase the number of parameters from $2$ to more than $3,600$. 
%
%
To calibrate a model with such a huge number of parameters one needs many more experimental data than we have access to right now. However, we tested the sensitivity of our model to variation in these parameters by repeating our simulations for 100 random sets of nutrient kinetic uptake and byproduct release rates ($\lambda$'s in our model), and found that this did not qualitatively change our central result (i.e., that the human gut microbiome is composed of roughly $N_\ell=4$ trophic levels with a byproduct fraction $f=0.9$). Surprisingly, our metabolome predictions were also relatively insensitive with respect to varying these parameters (Figure S1). The exact nature of the robustness of these metabolome predictions is beyond the scope of this paper, and the subject of future work.

\section*{Methods}
\subsection*{Obtaining data for microbial metabolic capabilities}
For information about the metabolic capabilities of human gut microbes, we adopted a recently published manually-curated database, NJS16, which includes such data for 570 common gut microbial species and 244 relevant metabolites from Ref. \cite{sung2017global}. This database recorded, for each microbial species, which metabolites each of the species could consume, and which they secreted as byproducts. Since we were interested in those metabolites that could be used for microbial growth, we removed metabolites such as ions (e.g., $\text{Na}^+$, $\text{Ca}^+$) from NJS16. Moreover, we constrained our analyses to microbes only, and therefore removed the 3 types of human cells from NJS16.
This left us with a database with 567 microbes, 235 metabolites and 4,248 interactions connecting these microbes with corresponding metabolites (see table S1 for the complete table of interactions). 

\subsection*{Obtaining metagenomic and metabolomic data}
To calibrate the key parameters of our model, we used a previously published dataset, namely a 16S rRNA sequencing study of 41 human individuals from rural and urban areas in Thailand \cite{kisuse2018urban}. From these data, we collected the reported 16S rRNA OTU abundances as well as their corresponding taxonomy. We explicitly removed all OTUs that did not have an assigned species-level taxonomy. The remaining OTUs explained roughly $71\% (\pm15\%)$ of the bacterial abundances per sample.

We then mapped these species names to those listed in the NJS16 database. We found an exact match for 110 species out of 208 in this table. In order to improve the species coverage from the abundance data, we manually mapped the remaining species in the following manner. For those genera in NJS16, whose member species had identical metabolic capabilities, we assumed that the capabilities of other, unmapped species from these genera were the same as these species. For several well-studied bacterial genera, such as \emph{Bacteroides}, we 
determined a 
``core'' set of metabolic capabilities (i.e., those metabolites that could either be consumed or secreted by all species in that genus), and assigned them to all unmapped species in that genus (i.e., those with known abundances, but otherwise understudied metabolic capabilities in NJS16). This allowed us to map an additional 20 microbial species from the abundance data, and incorporate into our model. Note that we did this additional mapping, only for those genera, where species metabolic capabilities were identical.

To quantify the metabolome levels in each individual, we used the available quantitative metabolome profiles (obtained via from CE-TOF MS) corresponding to the 41 individuals whose metagenomic samples we had. Here, we mapped the reported metabolites to our database of metabolic capabilities using KEGG identifiers, which revealed 84 such measured metabolites.

To make predictions about metabolic flow and effective diversity from our model, we used additional metagenomic datasets, namely those from the Human Microbiome Project (HMP) \cite{HMP2012} and MetaHIT \cite{Qin2010, qin2012metagenome}, for which we had microbial abundances, but no fecal metabolome. This resulted in an additional 380 human individuals, for which we obtained tables of MetaPhlAn2 microbial abundances, and mapped species names to those in NJS16 using the same procedure described above. Here, out of a total of 532 microbial species detected over these data, we could map and incorporate 316 species. Of these, 207 were mapped through an exact taxonomic match, and 109 by a genus-capability match. These incorporated species covered, on average, $90\%$ of the total microbial abundance in each individual sample.

\subsection*{Determining the components of the nutrient intake to the gut}
The inputs of our model are the \textcolor{black}{experimentally measured} relative abundances of microbial species in each individual, which are known (and described above), and the levels of various nutrients reaching their lower gut, which we fit using the model. \textcolor{black}{Note that we always used the experimentally measured relative microbial abundances, which simplified calculations and made the model easy to run. This also removed the model's dependence on the initial relative abundances, and the need for a new set of parameters to represent them. Moreover, this assumption is valid and self-consistent; our model's calculated abundances are very close to the experimentally observed abundances (see figure \ref{fig2}B). This is discussed in greater detail in the next section.} For simplicity, we did not explicitly include the various polysaccharides (dietary fibers, starch, etc.) known to constitute the bulk of an individual's diet. 
Instead, we chose not to include the polysaccharides themselves, but instead use their breakdown products as the direct nutrient intake to the gut. The reason for this is our limited quantitative understanding of the processes by which these polysachharides are converted to these breakdown products, e.g., the levels of extracellular enzymes, variability in their composition (their lability), etc. This curated nutrient intake consisted of 19 metabolites, such as arabinose, raffinose, and xylose (see table S2 for the complete list of metabolites). 

{\color{black}
\subsection*{Constructing and validating the trophic model}

Our model incorporates a set of observed microbial species abundances and the known metabolic cross-feeding interactions between these species, to calculate and predict both the step-wise metabolic flow through the lower gut, and the resulting fecal metabolome. The model does this on an individual-to-individual basis. We started simulating the model with the various levels of nutrients entering the lower gut, represented by the 19-dimensional vector $\vec{c}_{\text{nut}}$. Each element of $\vec{c}_{nut}$, say $c_{nut, i}$ represents the amount of one of the 19 metabolites entering the lower gut of that individual. We inferred these amounts through a fitting procedure described in the next section. Throughout this description, we use the subscript $i$ to refer to metabolites, and $\alpha$ to refer to microbial species. 

In the first trophic level, we calculated how these nutrients entering the gut were consumed by the gut microbiome and converted to microbial biomass, $\vec{B}$ and metabolic byproducts, $\vec{c}_{\text{layer}}$. For this, we calculated the relative increase in microbial biomass for each species, $\alpha$, as follows:

\begin{equation}
    B_\alpha = (1 - f) \cdot A_{in} \cdot \vec{c}_{nut},
\end{equation}
where $(1 - f)$ represents the fraction of consumed metabolites converted to biomass, and $f$ represents the fraction of input nutrients converted to metabolic byproducts. $A_{in}$ is a matrix which represents how each species takes up and consumes the nutrients it is capable of. Each term of this matrix, $(A_{in})_{\alpha,i}$ was set to zero if species $\alpha$ was incapable of consuming metabolite $i$ as a nutrient (using the set of microbial metabolic capabilities in table S1). If species $\alpha$ was instead capable of consuming metabolite $i$ as a nutrient, then $(A_{in})_{\alpha,i}$ was set as follows:

\begin{equation}
    (A_{in})_{\alpha,i} = \kappa_{i} \lambda_{\alpha,i} B_\alpha^{\text{exp}}.
\end{equation}

Here, $\lambda_{\alpha,i}$ represents the rate at which species $\alpha$ takes up nutrient $i$, $B_\alpha^{\text{exp}}$ is the experimentally measured abundance of strain $\alpha$, and $\kappa_i$ is a 
constant to normalize the relative microbial abundances of species capable of consuming nutrient $i$ to one. Throughout the manuscript, we set $\lambda_{\alpha,i}=1$ for all values of  $\alpha$ and $i$; this is because we lacked knowledge of the precise rates at which each species takes up different nutrients, and had insufficient data about microbial growth to  fit them using our model. To verify that this assumption did not significantly affect the predictions of our model, we repeated our metabolome predictions 100 times by assigning each value of $\lambda_{\alpha,i}$ randomly, chosen from a uniform distribution between 0 and 1 (see figure \ref{suppfig1}).

After calculating the contribution of nutrient consumption to microbial biomass, we computed the relative levels of the first level of metabolic byproducts produced by them, as follows:

\begin{equation}
    c_{1, i} = f A_{out} A_{in} \cdot \vec{c}_{nut},
\end{equation}
where the $1$ indicates that we were calculating the first layer of byproducts, and $i$, each metabolite which could be secreted as a byproduct. $A_{out}$ is  matrix which represents which byproduct each species could secrete, and in what amount. Each term of this matrix $(A_{out})_{i,\alpha}$ was set to zero if species $\alpha$ could not secrete metabolite $i$ as a byproduct (using the interactions in NJS16 described previously; see table S1). If species $\alpha$ was instead capable of secreting metabolite $i$ as a byproduct, then $(A_{out})_{i,\alpha}$ was set as follows:

\begin{equation}
    (A_{out})_{i,\alpha} = \frac{1}{(\mathcal{N}_{out})_\alpha},
\end{equation}
where $(\mathcal{N}_{out})_\alpha$ is the number of byproducts that species $\alpha$ was capable of secreting.

In the second trophic level (and all subsequent levels), we calculated how the byproducts secreted by the microbes in the previous step were consumed by the gut microbiome and converted to further biomass and byproducts. After $N_\ell$ such steps, we calculated the final microbial abundances, $\vec{B}$, and the accumulated metabolic byproducts, $\vec{c}_{\text{metabolome}}$. We would later compare these with the individual's experimentally measured metagenome and fecal metabolome, respectively. The final microbial abundances, $\vec{B}$, were calculated as follows:

\begin{equation}
    B_\alpha = \sum_{\ell=1}^{N_\ell} (1 - f)^\ell \cdot f^{\ell-1} \cdot A_{in} \cdot (A_{out} A_{in})^{\ell - 1} \cdot \vec{c}_{nut}.
\end{equation}

Here, we chose the appropriate number of trophic levels, $N_\ell$ and the byproduct fraction, $f$, by comparing the model's predicted fecal metabolome with the individual's experimentally measured metabolome. The number of levels and byproduct fraction that best matched the experimentally observed metabolomes, averaged over all individuals, were the ones that were considered to best represent their gut microbiome. To measure the best match, we used two different measures: (1) the Pearson correlation coefficient between the predicted and experimentally measured fecal metabolomes (see figure \ref{fig2}A), and (2) a logarithmic accuracy, i.e., the average difference between the log-transformed predicted and observed metabolome levels (see figure \ref{suppfig2}), i.e., $\frac{1}{19} \sum_{i=1}^{19} \left| \log_{10}(p_i) - \log_{10}(m_i) \right|$, where $m_i$ is the experimentally measured metabolome level of metabolite $i$, and $p_i$ is the predicted metabolome level of metabolite $i$, calculated by summing up the levels of all unused metabolites. Specifically, at each level, we calculated the byproducts similar to the first level (see equation (3)), as follows:

\begin{equation}
     \vec{c}_{\ell} = f^\ell (A_{out} A_{in})^{\ell-1} \cdot \vec{c}_{nut}.
\end{equation}

We split the byproducts at each level, $\vec{c}_{\ell}$, into two parts: a consumable part, $\vec{c}_{\ell}^{\text{ con}}$ and an unconsumable part, $\vec{c}_{\ell}^{\text{ uncon}}$. While the consumable part of the byproducts was available to the next trophic level of microbial species, the unconsumable part was composed of all the byproducts which could not be consumed by any microbial species in the individual's gut microbiome (i.e., it satisfied $A_{in} \vec{c}_{\ell}^{\text{ uncon}} = \vec{0}$). The former, consumable part was obtained by subtracting the unconsumable part from the generated byproducts at each level, i.e.,  $\vec{c}_{\ell}^{\text{ con}} = \vec{c}_{\ell} - \vec{c}_{\ell}^{\text{ uncon}}$. Finally, we calculated the predicted metabolome, $\vec{M}$, by adding up  the unconsumable byproducts from all previous levels with all the byproducts from the final trophic level, as follows:

\begin{equation}
     \vec{M} = \vec{c}_{N_\ell} + \sum_{\ell=1}^{N_\ell-1} \vec{c}_{\ell}^{\text{ uncon}}.
\end{equation}

Note that while the Pearson correlation (and its associated $P$-value) give an indication of the similarity in the trends predicted by our model with the experimentally observed metabolome, the logarithmic accuracy actually calculates the average error (measured in orders of magnitude) between the predicted and experimentally observed metabolomes. In both cases, we used the log-transformed values because we were interested in comparing the quality of our predictions with the experimental measurements at the level of resolution of an order of magnitude. This avoided overfitting in the model. Moreover, note that the nutrient input to the model (which we fit; see next section) resulted in a predicted set of microbial abundances, $\vec{B}$ (obtained from equation (5)) that were very close to the experimentally observed abundances. This allowed us to simplify our calculation; we used the experimentally measured microbial abundances instead of a more complicated, step-wise calculation in the sum of equation (5).

For each metabolome correlation coefficient that we calculated, we also corrected its associated $P$-value, in order to account for the two adjustable parameters in our model. 
We did this by adjusting (1) $t$-statistic: the adjusted $t$-statistic is obtained by dividing the original $t$-statistic by $\sqrt{\frac{n - 2 - p}{n - 2}}$, where $n$ was the number of metabolites (or points) that were used to measure the correlation, and $p$ was the number of adjustable model parameters (in our case, $p=2$); (2) $t$-test: typically the one-tailed t-test with degree of freedom $n-2$ is used to compute of $P$-value for the Pearson correlation coefficient. Here the one-tailed t-test with degree of freedom $n-2-p$ is used to account for adjustable model parameters.

}
\subsection*{Fitting and inferring the nutrient intake to the gut}
Simulating the model required us to know the nutrient intake to the gut, for which there are no available experimental measurements. Therefore, we inferred the amounts of these 19 intake metabolites by fitting the microbial abundances predicted by our model with those measured from each individual's microbiome. We used a nonlinear optimization technique for this (implemented as \texttt{lsqnonlin} in MATLAB R2018a, Mathworks Inc.). \textcolor{black}{We initially chose a random set of nutrient inputs, each chosen randomly from a uniform distribution between 0 and 1, and normalized so that all nutrient inputs summed up to one. For this random set of nutrient inputs, we calculated the predicted microbial abundances using equation (5). We then calculated the error in this prediction, by using the log-transformed differences between the predicted and experimentally measured microbial abundances, i.e.,  $\frac{1}{S} \sum_{\alpha=1}^{S} \left| log_{10}(p_\alpha) - log_{10}(m_\alpha) \right|$, where $S$ is the number of microbial species with non-zero abundances in the individual, $p_\alpha$ is the predicted relative abundance of species $\alpha$, and $m_\alpha$ is the experimentally measured abdunace of species $\alpha$. We then let the nonlinear optimization routine vary and choose that set of nutrient inputs, which minimized this error. We assumed that this set of nutrient inputs, which best explained the observed microbial abundances, given the microbial cross-feeding interactions, as the nutrient intake to the lower gut, or first trophic level, of that individual. Note that this is only step where we perform fitting in the model. All other subsequent steps, especially the prediction of the fecal metabolome, is an independent prediction from the model.} Typically, we fit 19 metabolite amounts for each human individual, who had roughly 80 microbial species.

{\color{black}
\subsection*{Shuffling microbial metabolic capabilities to test model predictions}
To test how good our model's gut metagenome and fecal metabolome predictions were against a null, or random, expectation, we repeated our simulations using a randomly shuffled set of microbial metabolic capabilities. For each individual microbial species, we picked one metabolite that they either could consume randomly, and swapped it with a metabolite that could be consumed by another microbial species. We also did this separately and independently with metabolites that they could secrete. Such swaps ensured that each microbial species could still consume and secrete the same number of metabolites as in the original dataset, but shuffled all the ecologically relevant metabolic relationships between species and metabolites. The swapping is performed three times as many the number of edges in the network to guarantee enough randomness. At the end of several rounds of swapping such relationships, we repeated our model's simulations exactly as described above, except with this shuffled set of microbial metabolic capabilities.
}

\subsection*{Calculating level-by-level diversity}
To quantify the diversity of microbes and metabolites at each trophic level across the 380 individuals we studied, we used three measures popular in the ecosystems literature: namely the $\alpha$-, $\beta$- and $\gamma$- diversity \cite{whittaker1972evolution, tuomisto2010diversity, tuomisto2010consistent}.
For each individual, we calculated the $\alpha$-diversity of microbes and metabolites on each of the trophic levels. For this we first quantified the relative contributions of a given level to microbial abundances, and separately to the fecal metabolome profile. The contribution of a given trophic level $\ell$ 
to the relative abundance of a species (microbial or, separately, metabolic) $i$ 
in a specific individual $j$ is given by $p_{i}(\ell, j)$ normalized by $\sum^S_{i=1}p_{i}(\ell, j)=1$.
The $\alpha$-diversity 
\begin{align*}
    D_{\alpha}(\ell) = \dfrac{1}{\langle \sum^S_{i=1}p_{i}(\ell, j)^2 \rangle_j},
\end{align*}
where 
$\langle \cdot \rangle_j$ represents taking the average across $380$ individuals used in our analysis.

Across all individuals, we calculated the $\gamma$-diversity of microbes and metabolites in their gut, which quantified the ``global'' diversity across all individuals, as:

\begin{align*}
    D_{\gamma}(\ell) = \dfrac{1}{{\sum^S_{i=1}p_{i}(\ell)^2}},   
\end{align*}
where $p_i(\ell)=\langle p_{i}(\ell, j) \rangle_j$ 
is the mean relative abundance of species (or metabolite) $i$ at the trophic level $\ell$ 
across all individuals used in our analysis.

Finally, to quantify the between-individual variability in microbial and metabolite diversity, we calculated the overall $\beta$-diversity, which is the ratio of the global to local diversity, as:

\begin{align*}
    D_{\beta}(\ell) = \frac{D_{\gamma}(\ell)}{D_{\alpha}(\ell)}.
\end{align*}


\subsection*{Code availability}
All computer code and extracted data files used in this study are available at the following URL: \url{https://github.com/eltanin4/trophic_gut}.

\section*{Supplementary Figures and Tables}



\paragraph*{Figure S1}
\label{S3_Fig}
{\bf Effect of changing kinetic parameters on model prediction.} Scatter plot of the measured and predicted metabolome where, instead of considering equal specific nutrient uptake and byproduct release rates, $\lambda$'s in our model, we take several random sets (in black). Error bars (in black) indicate standard deviation in the predicted levels of specific metabolites for different sets of $\lambda$'s. The solid line represents $x=y$. Red squares indicate the predicted metabolome for the default set of kinetic parameters used, i.e., when all of $\lambda$'s were set equal to 1.

\paragraph*{Table S1}
\label{S1_Table}
{\bf Microbial and metabolite interactions used in the model.} Table of all 4,248 interactions between microbes and metabolites used in the model, from Ref. \cite{sung2017global}. 

\paragraph*{Table S2}
{\bf Components of the nutrient intake to the gut.} List of all 19 metabolites used to fit the gut nutrient intake in the model.

\paragraph*{Table S3}
{\bf Metabolome predictions of the model for 380 individuals from the Human Microbiome Project (HMP) and the MetaHIT study.} All metabolites in metabolome predicted by the model with global parameters $f = 0.9$ and $N_{\ell}=4$ for all 380 individuals are listed.

\section*{Acknowledgments}
A.G. acknowledges support from the Simons Foundation and the American Physical Society. We thank Parth Pratim Pandey for useful discussions.

\section*{Conflicts of interest}
The authors declare that there are no competing interests.

\nolinenumbers
\bibliographystyle{plos2015}
\bibliography{trophic_model}

\clearpage
\section*{Supplementary Figures}
\renewcommand{\thefigure}{S\arabic{figure}}
\setcounter{figure}{0}



\vfill
\begin{figure}[h]
\centering
\includegraphics[width=0.5\linewidth]{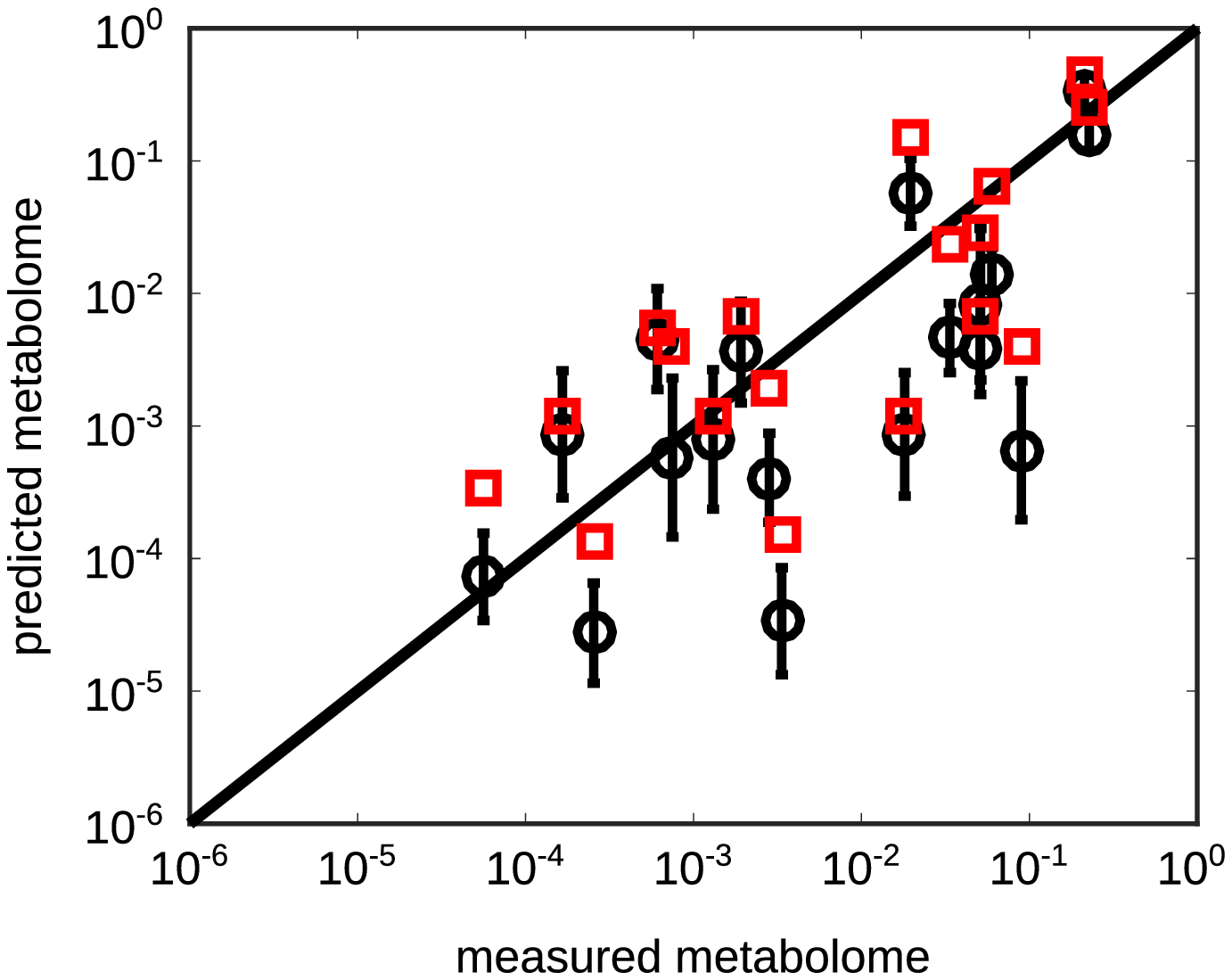}
\caption{{\bf Effect of changing kinetic parameters on model prediction.}
 Scatter plot of the measured and predicted metabolome where, instead of considering equal specific nutrient uptake and byproduct release rates, $\lambda$'s in our model, we take several random sets (in black). Error bars (in black) indicate standard deviation in the predicted levels of specific metabolites for different sets of $\lambda$'s. The solid line represents $x=y$. Red squares indicate the predicted metabolome for the default set of kinetic parameters used, i.e., when all of $\lambda$'s were set equal to 1.}
\label{suppfig1}
\end{figure}
\vfill

\vfill
\begin{figure}[h]
\centering
\includegraphics[width=0.6\linewidth]{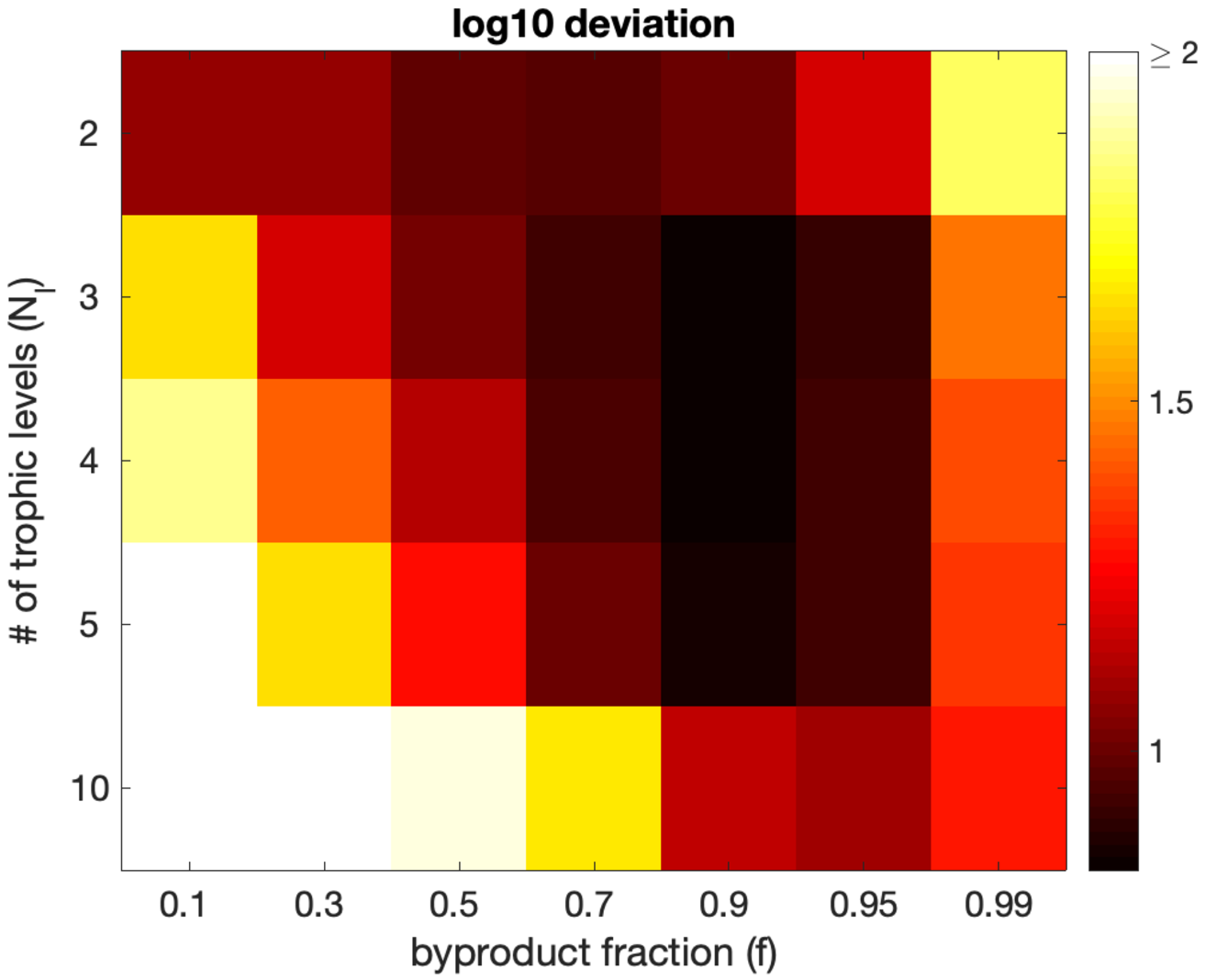}
\caption{{\bf Calibrating the model parameters using logarithmic accuracy.}
Heatmap of the logarithmic accuracy between experimentally measured and predicted fecal metabolomes for different combinations of parameters $f$ and $N_{\ell}$. The logarithmic accuracy quantifies the average order-of-magnitude error in our predicted fecal metabolome, when compared with the experimentally measured one (see Methods for details). The plotted value is the logarithmic accuracy averaged over 41 individuals in Ref. [22]}
\label{suppfig2}
\end{figure}
\vfill

\vfill
\begin{figure}[h]
\centering
\includegraphics[width=0.8\linewidth]{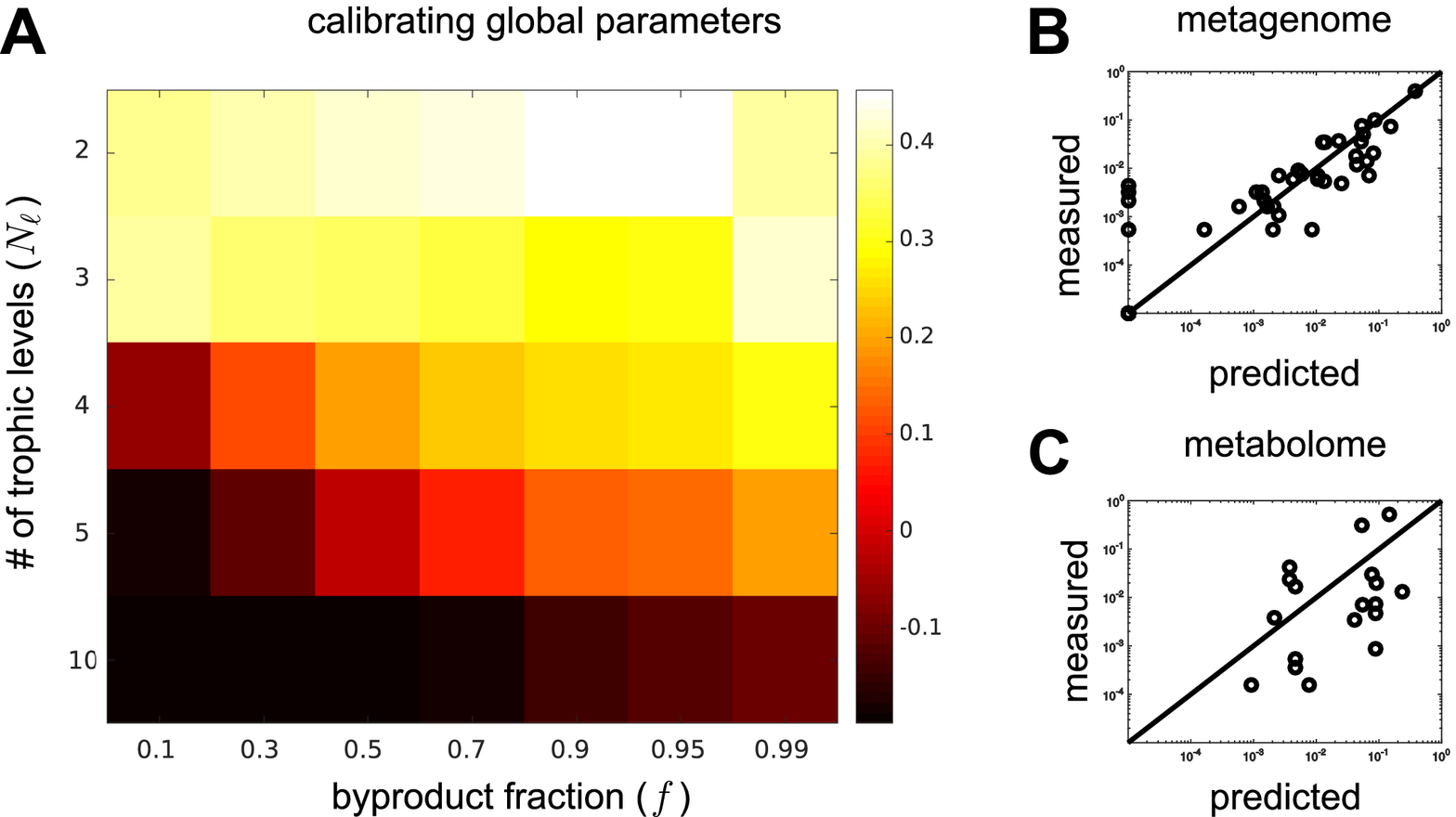}
\caption{{\bf Results for model calibration after shuffling microbial metabolic capabilities.}
Same as figure 2 from the main text, except that the simulations have been performed after shuffling the set of microbial metabolic capabilities, to test the dependence of our predictions on metabolic interactions (see Methods for details).
(A) Heatmap of the Pearson correlation between experimentally measured and predicted metabolomes for different combinations of parameters $f$ and $N_\ell$. The plotted value is the correlation coefficient averaged over 41 individuals in Ref. \cite{kisuse2018urban}. For this shuffled network, the best average Pearson correlation coefficient 0.44 is given by $f=0.9$ and $N_{\ell}=2$. Panel (B) and (C) are generated by those global parameters. (B) Comparison between 
the experimentally observed bacterial abundances in a representative individual (y-axis) and their best fits from our model (x-axis) with 
$f=0.9$ and $N_\ell=2$. (C) Comparison between the experimentally observed fecal metabolome (y-axis) and the predictions of our model (x-axis) with 
$f=0.9$ and $N_\ell=2$ in the same individual shown in panel (B) (Pearson correlation $0.32$; $P$-value $0.19$).}
\label{suppfig3}
\end{figure}
\vfill

\vfill
\begin{figure}[h]
\centering
\includegraphics[width=0.5\linewidth]{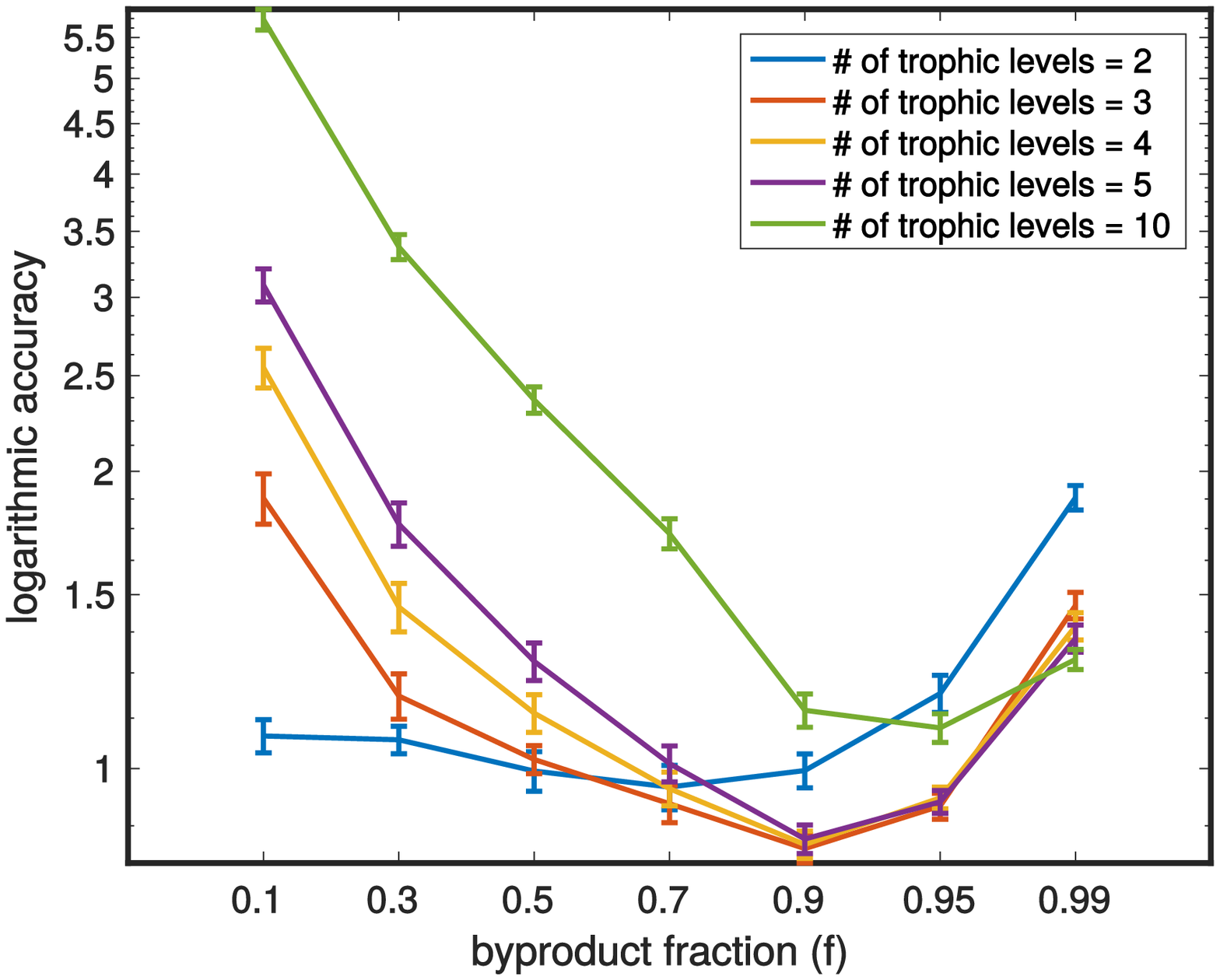}
\caption{{\bf Testing model predictions for byproduct fractions beyond 0.9}
Logarithmic accuracy of the model's predictions between the metabolomes (see Methods for details) for byproduct fraction, $f$, values 0.95 and 0.99. Higher values on the y-axis means worse predictions. This suggests that at 4 trophic levels, $f=0.9$ gives the best calibration.}
\label{suppfig4}
\end{figure}
\vfill

\vfill
\begin{figure}[h]
\centering
\includegraphics[width=0.5\linewidth]{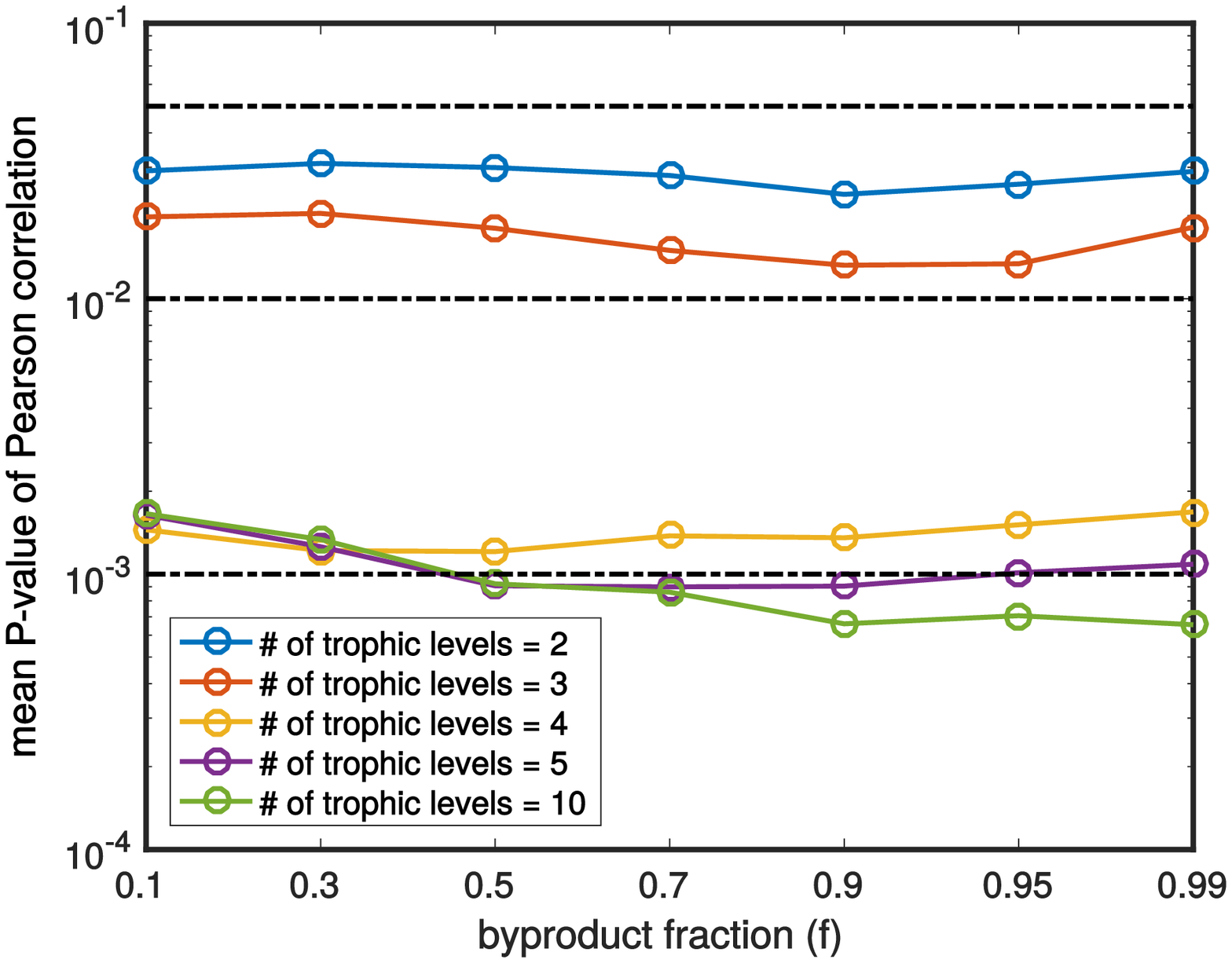}
\caption{{\bf Adjusted $P$-values for the model predictions}
Blue nodes correspond to metabolites, red nodes to the microbes. Blue edges show metabolite consumption, red - production.}
\label{suppfig5}
\end{figure}
\vfill

\clearpage

\vfill
\begin{figure}[h]
\centering
\includegraphics[width=0.9\linewidth]{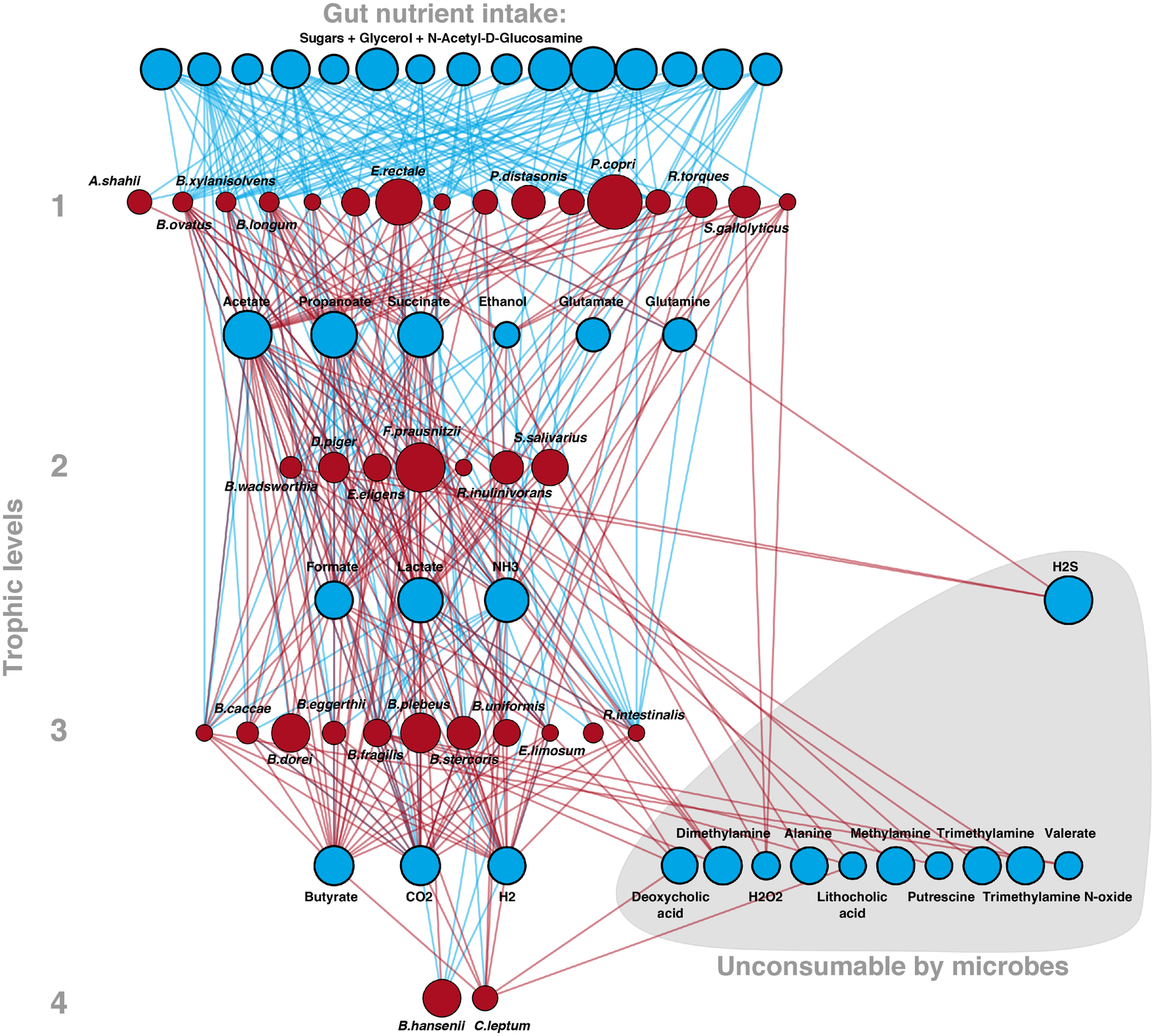}
\caption{{\bf Layer-wise network for one of the individuals from the calibrated dataset.}
Blue nodes correspond to metabolites, red nodes to the microbes as in figure \ref{fig1}B. Blue edges show metabolite consumption, red - production.}
\label{suppfig6}
\end{figure}
\vfill

\end{document}